\documentclass{iopconfser}
\usepackage{graphicx}	
\usepackage{amsmath}	
\usepackage{amssymb}	
\usepackage{siunitx}
\usepackage{natbib}

\usepackage{aas_macros}

\begin{document}

\title{Mode energy partition in partially ionized compressible MHD turbulence}

\author{Yue Hu$^{1,2}$}

\affil{$^1$Institute for Advanced Study, 1 Einstein Drive, Princeton, NJ 08540, USA}
\affil{$^2$NASA Hubble Fellow}

\email{yuehu@ias.edu}

\begin{abstract}
We investigate how neutral-ion collisional damping modifies the spectral properties and energy partition of compressible MHD turbulence using a suite of 3D two-fluid simulations. By systematically varying the neutral-ion coupling strength and decomposing the turbulent velocity field into Alfv\'en, slow, and fast  (polarization) modes, we quantify how each mode responds to the transition from strong to weak coupling. In the strong-coupling regime, the Alfv\'en and slow modes follow nearly Kolmogorov $k^{-5/3}$ spectra and dominate the kinetic energy budget, while fast modes exhibit a steeper spectrum and contribute $\sim$10\% of the total energy. As the coupling weakens and neutral-ion damping becomes significant, all mode spectra steepen, approaching a dissipation-dominated $k^{-4}$ spectrum, except that the slope mode's spectrum parallel to the mean magnetic field has a power-law slope shallower than -4. While the total kinetic energy is reduced in the weak coupling regime, the slow-mode energy fraction increases substantially toward small scales, whereas the Alfv\'en-mode fraction decreases correspondingly. In contrast, the fast-mode energy fraction remains largely insensitive to coupling strength. These results demonstrate that partial ionization not only steepens the turbulent spectra but also reshapes the mode energy distribution, enhancing the relative importance of the slow mode while suppressing Alfv\'en mode in the damping regime. Our findings have important implications for turbulence-driven processes in the partially ionized interstellar medium, including cosmic-ray transport and acceleration.
\end{abstract}

\section{Introduction}
Magnetohydrodynamic (MHD) turbulence is pervasive throughout the interstellar medium (ISM), extending from tens-of-parsec molecular clouds to sub-parsec dense clumps and even AU-scale structures in the very local ISM \citep{1981MNRAS.194..809L,1995ApJ...443..209A,2010ApJ...710..853C,2022ApJ...934....7H}. It plays a key role in shaping the dynamics and thermodynamics of the ISM \citep{1998JGR...103.1889L, 2000ApJ...540..271V,2003PhPl...10.1954N, 2008A&A...486L..43H,2020ApJ...905..129H,2022MNRAS.510.4952L,2025ApJ...988..188H,2025ApJ...986...62H}, regulating star formation \citep{2004RvMP...76..125M, 2007ARA&A..45..565M,2012nsf....1211729M,2012ApJ...761..156F}, and governing the transport and acceleration of cosmic rays \citep{1966ApJ...146..480J, 2002PhRvL..89B1102Y, 2008ApJ...673..942Y, 2013ApJ...779..140X,2023FrASS..1054760L,2024NatCo..15.1006H,2025arXiv250507421H,2025arXiv250907104H}. MHD turbulence is thus a fundamental bridge between small-scale physical processes and the large-scale evolution of galaxies.

Compressible MHD turbulence can be decomposed into three eigenmodes—fast, slow, and Alfv\'en—which provide a natural framework for characterizing the fundamental properties of turbulent fluctuations. \citet{2003MNRAS.345..325C} introduced a mode decomposition method and analyzed the scaling and anisotropy associated with each mode. Their results showed that the Alfv\'en and slow modes dominate the kinetic energy budget and follow a Kolmogorov-like $k^{-5/3}$ spectrum with scale-dependent anisotropy \citep{LV99}, whereas fast modes typically contribute only $\sim$10--20\% of the total kinetic energy and exhibit an isotropic $k^{-2}$ spectrum under supersonic conditions \citep{2003MNRAS.345..325C,2010ApJ...720..742K,2022MNRAS.512.2111H,2025ApJ...992L..28H}.

However, these studies assumed fully ionized turbulence or relied on single-fluid MHD approximations. In many ISM environments—such as cold molecular clouds—the gas is only weakly ionized, giving rise to additional physical processes including ion-neutral collisions and the accompanying collisional damping of turbulence \citep[e.g.,][]{1986MNRAS.220..133D,1996ApJ...465..775B,2011MNRAS.415.3681T,2015ApJ...810...44X}. These effects can significantly modify both the local turbulent dynamics \citep{2010MNRAS.406.1201T,2010ApJ...720.1612M,2015ApJ...805..118B,2024MNRAS.527.3945H} and potentially the distribution of energy among the modes. Consequently, a two-fluid (ion + neutral) description is essential for capturing the relevant physics in partially ionized regions of the ISM and for understanding the fundamental behavior of MHD turbulence under realistic interstellar conditions. 

In this work, we analyze 3D two-fluid simulations of MHD turbulence \citep{2024MNRAS.527.3945H}, systematically varying the ion–neutral coupling strength to probe different regimes of their interaction. By decomposing the velocity fluctuations into the three MHD modes, we aim to determine how collisional ion–neutral damping modifies their spectral properties and distributes energy among the Alfv\'en, slow, and fast modes.

\section{Methodology}
\subsection{Two-fluid simulations}

The 3D two-fluid simulations analyzed in this work are performed using the AthenaK code \citep{2024arXiv240916053S} and have previously been employed to study the damping of MHD turbulence \citep{2024MNRAS.527.3945H}. The simulated magnetofluid consists of an ionized component (ions plus electrons) and a neutral component, evolving under the assumptions of an isothermal equation of state and periodic boundary conditions. The computational domain is discretized using a uniform grid of $480^3$ staggered cells. We briefly summarize the key simulation parameters here; further details can be found in \cite{2024MNRAS.527.3945H}.

\subsubsection{Initial conditions}

Initially, both ion and neutral densities are spatially uniform, and the magnetic field is uniform and oriented along the $z$-axis. The ionization fraction is set to $\xi_i = \frac{\rho_i}{\rho_i + \rho_n} = 0.1$, where $\rho_i$ and $\rho_n$ denote the ion and neutral mass densities, respectively. Turbulence is driven solenoidally at a peak wavenumber of $k=2$, with numerical dissipation occurring at scales of approximately 10 grid cells. To drive turbulent motions in ions and neutrals, a stochastic forcing term is applied to both. Explicitly, the forcing terms are weighted by ion and neutral densities to achieve the same injected turbulent velocities in the two fluids.

The scale-free turbulence is characterized by the sonic Mach number, $M_s = v_{\rm inj}/c_s$, and the Alfv\'en Mach number, $M_A = v_{\rm inj}/v_A$, where $v_{\rm inj}$ is the turbulence injection velocity and $v_A$ is the total Alfv\'en speed calculated from total density, i.e., ion density plus neutral density. Throughout this work, we fix both $M_s$ and $M_A$ to be of order unity, ensuring that the simulations reside in the strong eddy-like turbulence regime. 

\subsubsection{Ion--neutral coupling}

The coupling between ions and neutrals is characterized by two collisional frequencies. The neutral-ion collision frequency is defined as $\nu_{ni} = \gamma_{\rm d} \rho_i$,
while the ion-neutral collision frequency is given by $\nu_{in} = \gamma_{\rm d} \rho_n$,
respectively \citep{1992pavi.book.....S}. Here $\gamma_{\rm d}$ represents the drag coefficient governing the momentum exchange between ions and neutrals through collisions (see \citealt{2024MNRAS.527.3945H} for the two-fluid equations).

The typical ionization fraction in a molecular cloud is approximately $10^{-6}$. Simulating such a low ionization fraction directly is computationally prohibitive with current resources. Instead, to probe different coupling regimes, we vary the drag coefficient $\gamma_{\rm d}$. This allows us to transition the neutral--ion collision frequency from the strong-coupling regime (large $\nu_{ni}$), where ions and neutrals are tightly locked, to the weak-coupling regime (small $\nu_{ni}$), where the two fluids dynamically decouple. This approach is justified by the findings in the Appendix of \citet{2024MNRAS.527.3945H}, which demonstrated that the damping effects resulting from reducing $\gamma_{\rm d}$ are physically equivalent to those achieved by reducing the ion density $\rho_i$.

Furthermore, while a typical molecular cloud spans $\sim 10$~pc, ion-neutral decoupling occurs at much smaller scales, around $0.01$~pc. In our simulations with $\gamma_{\rm d}=25$, decoupling occurs at a normalized wavenumber near $k \sim 3$, suggesting that our simulation box effectively represents a physical scale slightly larger than, but comparable to, this $0.01$~pc decoupling scale. Assuming a cloud temperature of $10$~K (sound speed $c_s \approx 0.2$~km~s$^{-1}$) and a large-scale turbulent velocity of $\sim 2$~km~s$^{-1}$ at 10~pc, the turbulent cascade would reduce the velocity dispersion at the $0.01$~pc scale to approximately $0.2$~km~s$^{-1}$ (assuming Kolmogorov scaling). Consequently, $M_s$ at this scale is expected to be close to unity.


\subsection{Mode decomposition}
Compressible MHD turbulence can be decomposed into three distinct modes: the incompressible Alfv\'en mode and the compressible fast and slow modes. \citet{2003MNRAS.345..325C} introduced a decomposition method that separates these modes in Fourier space. The corresponding basis vectors \(\hat{\boldsymbol{\xi}}\) defining the mode decomposition are given by:
\begin{equation}
\label{eq:modedecomposition}
\begin{aligned}
\hat{\boldsymbol{\xi}}_{\rm s} &\propto \left(1+\frac{\beta}{2}-\sqrt{D}\right)k_{\bot}\hat{k}_{\bot} + \left(-1+\frac{\beta}{2}-\sqrt{D}\right)k_{\parallel}\hat{k}_{\parallel},\\
\hat{\boldsymbol{\xi}}_{\rm f} &\propto \left(1+\frac{\beta}{2}+\sqrt{D}\right)k_{\bot}\hat{k}_{\bot} + \left(-1+\frac{\beta}{2}+\sqrt{D}\right)k_{\parallel}\hat{k}_{\parallel}, \\
\hat{\boldsymbol{\xi}}_{\rm a} &\propto \hat{k}_{\parallel} \times \hat{k}_{\bot},
\end{aligned}
\end{equation}
where \(D = (1+\beta/2)^2 - 2\beta\cos^2\theta\), \(\beta = 2(M_A/M_s)^2\), and \(\theta\) is the angle between the wavevector \(\hat{\mathbf{k}}\) and the mean magnetic field \(\mathbf{B}_0\). Here the subscripts \(a\), \(f\), and \(s\) referring to the Alfv\'en, fast, and slow modes, respectively. By projecting the total velocity field onto these basis vectors in Fourier space and then performing an inverse Fourier transform, one obtains the velocity field associated with each mode. Note that in the weak-coupling regime, $M_A$ is calculated solely from the ion density. In this regime, the decomposition method \citep{2003MNRAS.345..325C} may not necessarily decompose the true two-fluid eigenmodes; rather, it may primarily serve as a polarization decomposition relative to the mean magnetic field. For simplicity, however, we retain the term `mode' throughout the paper.

\begin{figure*}[h]
\centering
\includegraphics[width=0.99\linewidth]{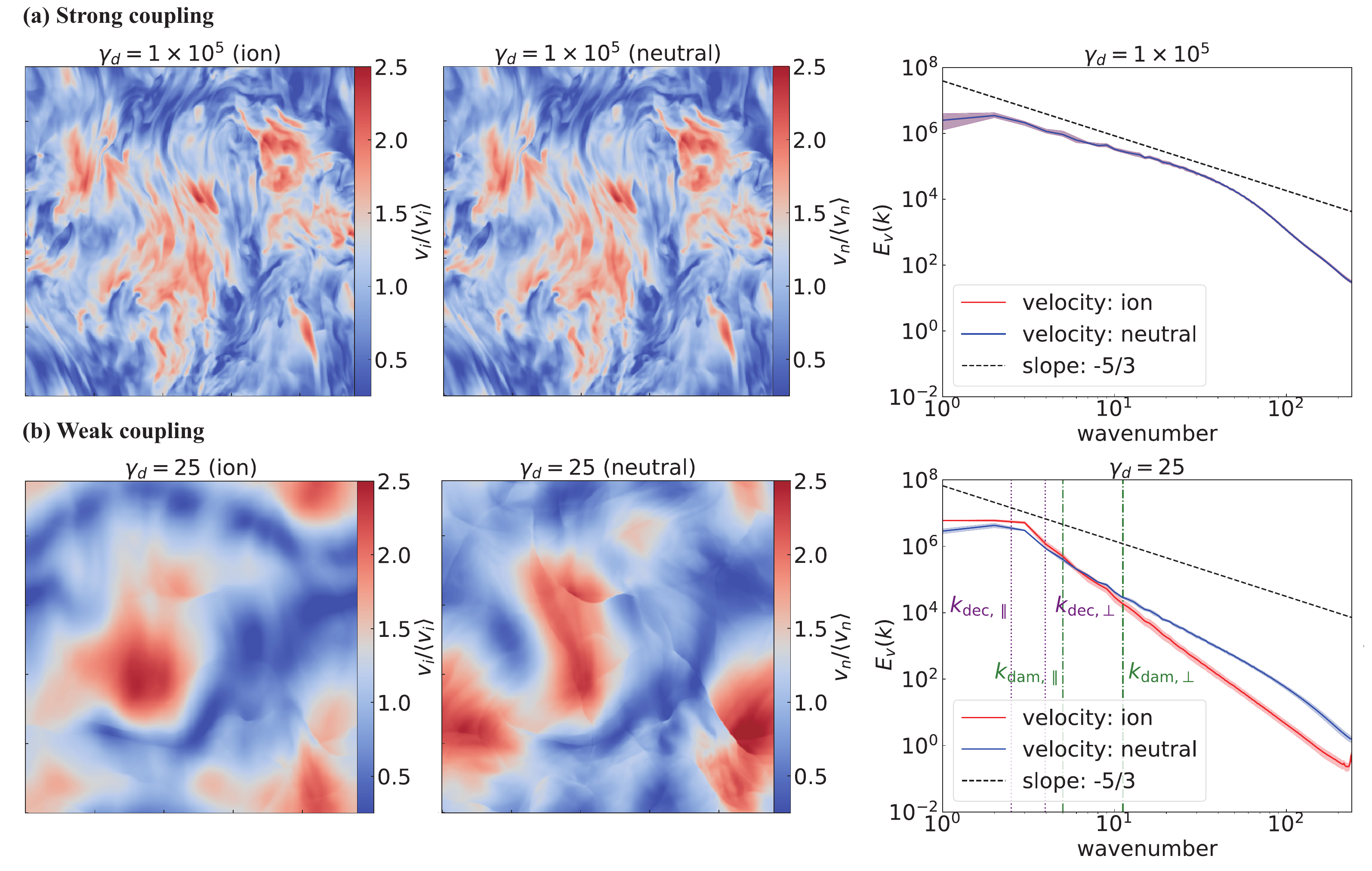}
        \caption{Left and middle panels: 2D slices of the ion (left) and neutral (middle) velocity fields, normalized by their mean amplitudes. The viewing direction is perpendicular to the mean magnetic field, which is oriented vertically. Right panels: Turbulent kinetic energy spectra of ions and neutrals. The shaded regions indicate temporal variations over several snapshots after the turbulence has reached a statistically steady state. The dotted lines show a $-5/3$ slope for reference to the Kolmogorov inertial-range scaling. The dash–dotted lines mark the theoretically predicted neutral–ion parallel decoupling and damping wavenumbers for Alfv\'enic MHD turbulence. These wavenumbers are not shown when they exceed the numerical dissipation scale, $k \sim 50$. Modified from \cite{2024MNRAS.527.3945H}.}
    \label{fig:vmap}
\end{figure*}

\section{Results}
\subsection{Neutral-ion collisional damping of MHD turbulence}
We present 2D slices of the ion and neutral velocity fields in Fig.~\ref{fig:vmap}, taken perpendicular to the mean magnetic field at the box center. We show two representative cases: strong coupling with $\gamma_{\rm d}=1\times10^5$ and weak coupling with $\gamma_{\rm d}=25$. The values of $\gamma_{\rm d}$ are expressed in numerical units; to obtain a dimensionless form, one may divide $\gamma_{\rm d}$ by $v_{\rm inj}/(L_{\rm inj}\rho_i)$, which is approximately 10 for this set of simulations.

In the strong-coupling case, the ion and neutral velocity structures are nearly identical. Their kinetic energy spectra both follow an approximate Kolmogorov scaling with a slope of $-5/3$. The parallel and perpendicular decoupling wavenumbers, $k_{\rm dec,\parallel}$ and $k_{\rm dec,\bot}$, exceed the numerical dissipation wavenumber ($\sim 50$), indicating that ions and neutrals remain well coupled across all resolved length scales.

In the weak-coupling case, where neutrals decouple from ions and ion–neutral collisional damping of MHD turbulence becomes important, the velocity structures of the two fluids differ markedly. Small-scale fluctuations are visibly damped in both components. The ion and neutral spectra also diverge starting from the perpendicular damping scale $k_{\rm dam,\bot}$ and the ion spectrum steepens to a slope of approximately $-3.2$, reflecting strong damping of the turbulent cascade, while the neutral spectrum remains shallower but still steeper than the Kolmogorov scaling.

\begin{figure*}[h]
\centering
\includegraphics[width=0.99\linewidth]{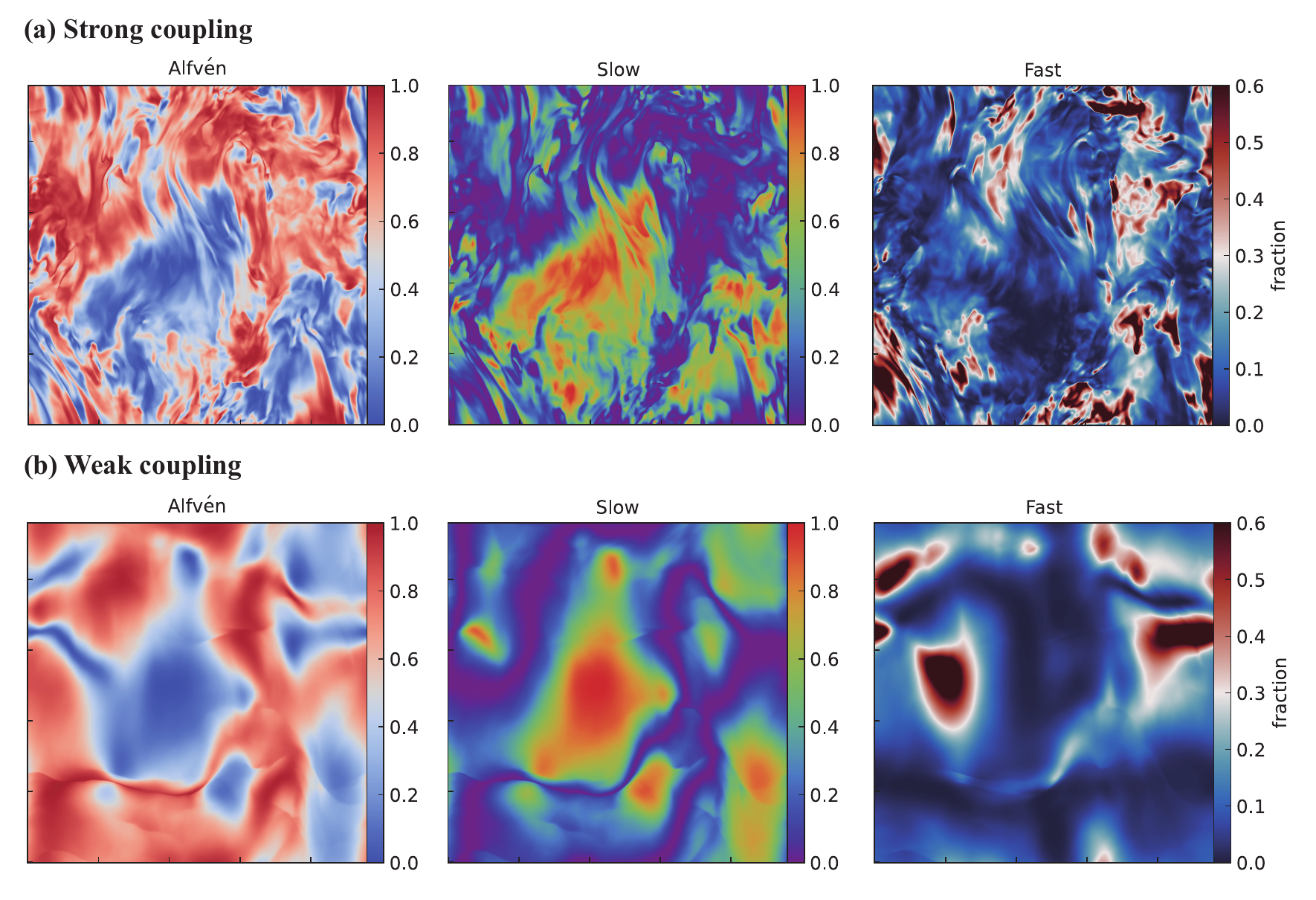}
        \caption{2D slices of the kinetic energy fraction in the decomposed MHD modes: Alfvén (left), slow (middle), and fast (right). Panel (a) shows the strong-coupling case with $\gamma_{\rm d}=10^5$, whereas panel (b) corresponds to the weak-coupling case with $\gamma_{\rm d}=25$.}
    \label{fig:mmap}
\end{figure*}

\begin{figure*}[p]
\centering
\includegraphics[width=0.99\linewidth]{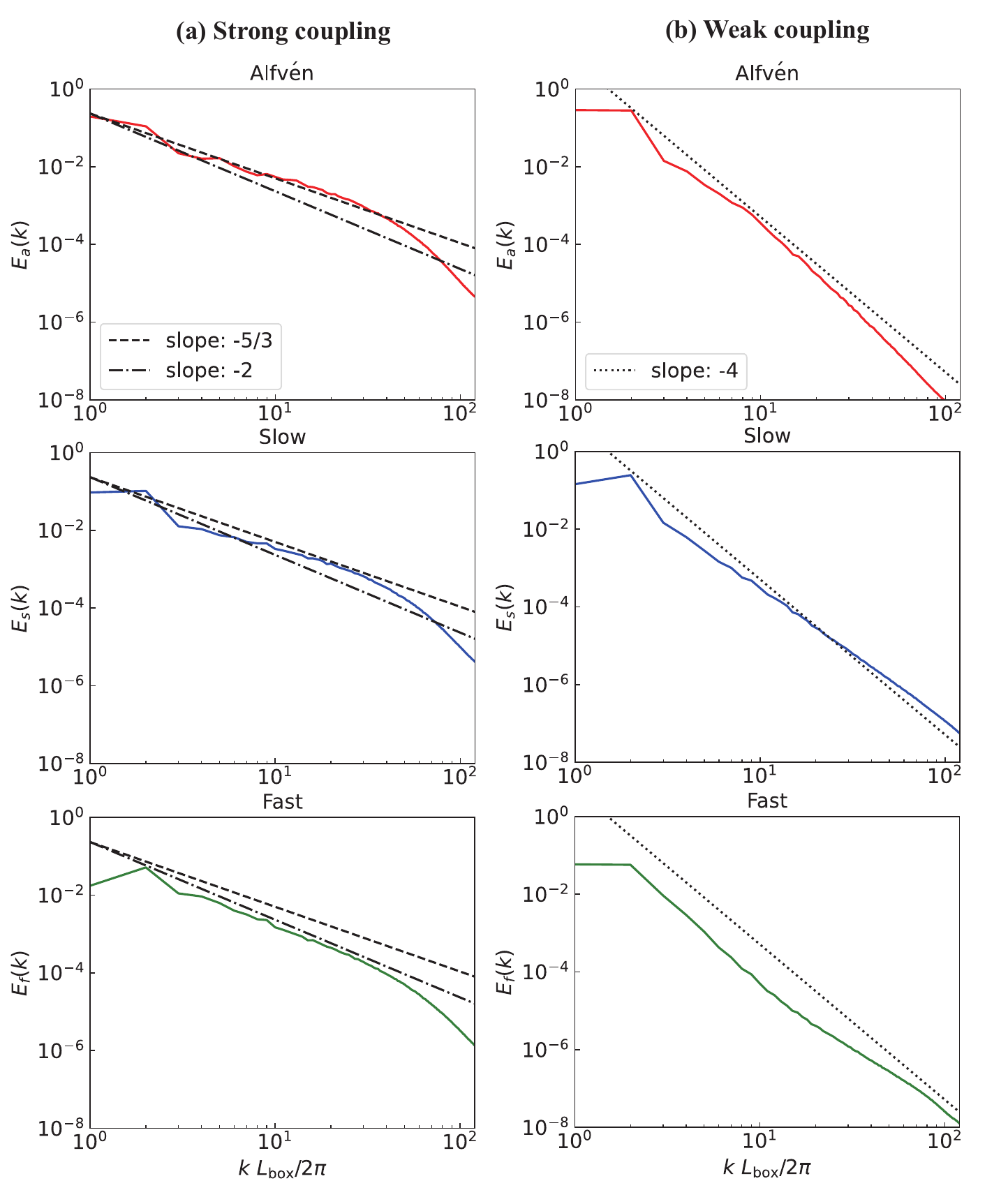}
        \caption{Turbulent kinetic energy spectra $E_i(k)$ of ions' decomposed MHD modes: Alfv\'en (top), slow (middle), and fast (bottom). The subscript $i = a, s, {\rm or}~f$ denotes the corresponding mode. Panel (a) shows the strong-coupling case with $\gamma_{\rm d}=10^5$, while panel (b) corresponds to the weak-coupling case with $\gamma_{\rm d}=25$. The dashed and dash–dotted lines indicate the Kolmogorov ($-5/3$) and Burgers ($-2$) scalings expected in the inertial range, while the dotted line with slope $-4$ illustrates the steepened spectrum in the dissipation-dominated regime caused by neutral-ion damping.}
    \label{fig:spectrum}
\end{figure*}

\subsection{Mode decomposition}
We perform mode decomposition of the ion velocity fields for both the strong-coupling ($\gamma_{\rm d}=1\times10^5$) and weak-coupling ($\gamma_{\rm d}=25$) regimes. Figure~\ref{fig:mmap} shows 2D slices of the Alfv\'en, slow, and fast modes' fractional contributions to the turbulent kinetic energy. The fraction is defined as $E_i/E$, where $E = E_a + E_s + E_f$ and  $E_i = \iiint \frac{1}{2} v_i^2 \, dx\,dy\,dz$,
with the subscript $i = a, s, \text{or}\ f$ denoting the corresponding mode. Here, density is ignored in the kinetic energy because density fluctuations are not significant in the transonic regime.

In both coupling regimes, the Alfv\'en mode dominates the global energy budget, followed by the slow mode, while the fast mode contributes the least. Locally, however, any of the three modes may dominate depending on the spatial location. For the fast mode in particular, regions with high local energy fraction tend to be spatially compact and concentrated on small scales.

\begin{figure*}[p]
\centering
\includegraphics[width=0.99\linewidth]{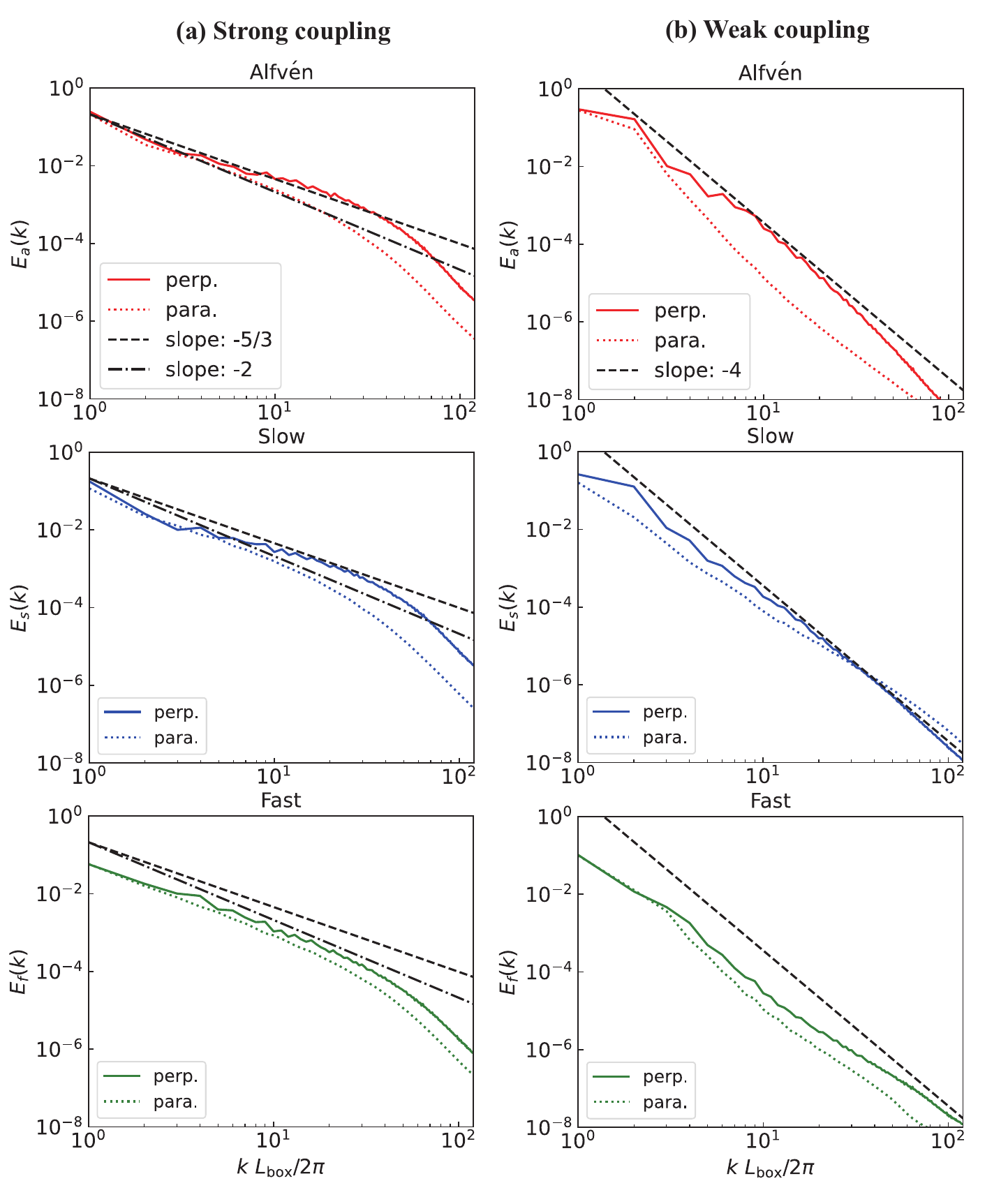}
        \caption{Turbulent kinetic energy spectra $E_i(k)$ of ions' decomposed MHD modes: Alfv\'en (top), slow (middle), and fast (bottom). The subscript $i = a, s, {\rm or}~f$ denotes the corresponding mode. The spectra are decomposed into parallel (para.) and perpendicular (perp.) components, with respect to the mean magnetic field. 
        Panel (a) shows the strong-coupling case with $\gamma_{\rm d}=10^5$, while panel (b) corresponds to the weak-coupling case with $\gamma_{\rm d}=25$. The dashed and dash–dotted lines indicate the Kolmogorov ($-5/3$) and Burgers ($-2$) scalings expected in the inertial range, while the dotted line with slope $-4$ illustrates the steepened spectrum in the dissipation-dominated regime caused by neutral-ion damping.}
    \label{fig:spectrum_decomp}
\end{figure*}

\subsection{Kinetic energy spectra of Alfv\'en, slow, and fast modes}
In Fig.~\ref{fig:spectrum}, we show the kinetic energy spectra of the Alfv\'en, slow, and fast modes in the strong-coupling ($\gamma_{\rm d}=1\times10^5$) and weak-coupling ($\gamma_{\rm d}=25$) regimes. 

In the strong-coupling case, the Alfv\'en-mode spectrum follows an approximate Kolmogorov scaling with a slope of $-5/3$, while the slow-mode spectrum is slightly shallower. The fast-mode spectrum is steeper than $-5/3$, approaching the characteristic $k^{-2}$ scaling. In the weak-coupling regime, where neutral–ion collisional damping becomes important, the Alfv\'en-mode spectrum steepens dramatically, approaching a slope slightly steeper than $-4$. This steepening reflects the dissipation-dominated regime induced by ion–neutral damping. The slow- and fast-mode spectra also exhibit pronounced steepening: the slow mode remains slightly shallower than $k^{-4}$, whereas the fast mode follows a $\sim k^{-4}$ slope up to $k\approx20$ before becoming somewhat shallower at higher wavenumbers. It indicates that the Alfv\'en mode's kinetic energy falls faster than the slow mode's kinetic energy. 

Likely, this different damping is fundamentally tied to ion–neutral drift and the degree to which neutrals are forced to participate in magnetically mediated motions. Alfv\'enic turbulence is predominantly transverse; consequently, when neutrals decouple, the ions experience significant drag, leading to efficient energy dissipation. In contrast, the field-parallel compressive motions associated with the slow mode may generate less ion–neutral drift, thereby reducing frictional dissipation \citep{2016ApJ...826..166X}. This interpretation is supported by the energy spectra decomposed into parallel and perpendicular components relative to the mean magnetic field, as shown in Fig.~\ref{fig:spectrum_decomp}. In the weak-coupling regime, the slow mode’s parallel spectrum exhibits a slope shallower than $-4$, whereas its perpendicular spectrum is close to $-4$. Notably, both the parallel and perpendicular spectra of the Alfv\'en mode exhibit steep slopes close to $-4$. This indicates that the slow mode’s parallel component dissipates less efficiently than both its perpendicular counterpart and the Alfv\'en mode in the weak-coupling limit.

\begin{figure*}[h]
\centering
\includegraphics[width=0.99\linewidth]{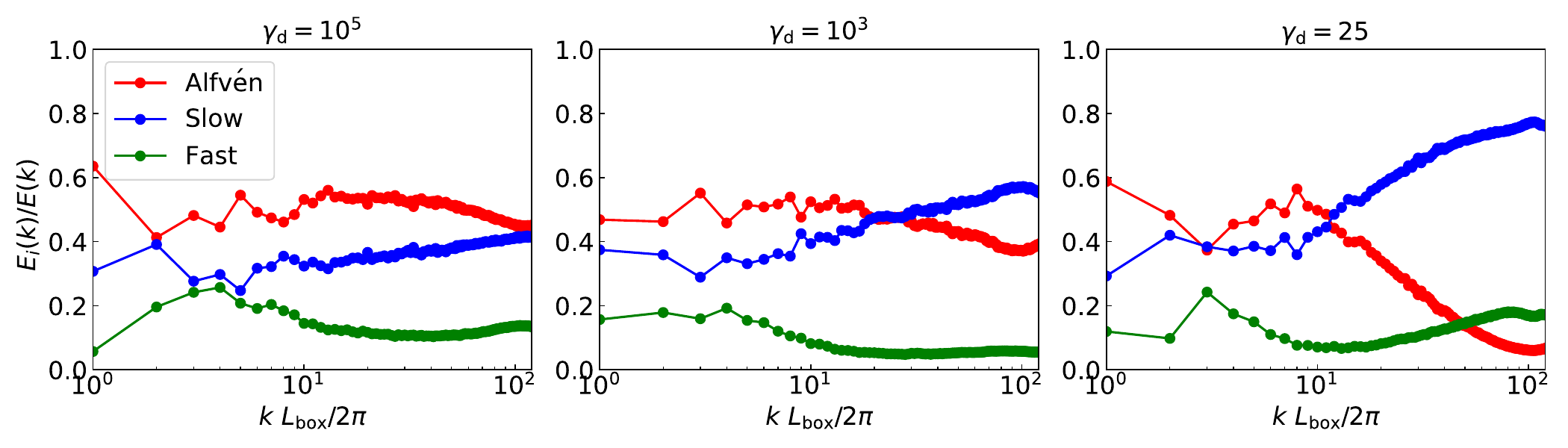}
        \caption{Fraction of turbulent kinetic energy in the Alfv\'en (red), slow (blue), and fast (green) modes as a function of wavenumber. Each fraction is defined as $E_i(k)/E(k)$, where $E(k)=E_a(k)+E_s(k)+E_f(k)$ and the subscript $i = a, s, {\rm or}~f$ denotes the corresponding mode.}
    \label{fig:fraction_k}
\end{figure*}

\subsection{Kinetic energy partition among Alfv\'en, slow, and fast modes}
In Fig.~\ref{fig:fraction_k}, we show the fraction of turbulent kinetic energy in the Alfv\'en, slow, and fast modes as a function of wavenumber. Each fraction is defined as $E_i(k)/E(k)$, where $E(k)=E_a(k)+E_s(k)+E_f(k)$ and the subscript $i=a, s,\text{ or }f$ denotes the corresponding mode. We include three coupling strengths in the calculation: $\gamma_{\rm d}=10^5$, $10^3$, and $25$.

In the strong-coupling case ($\gamma_{\rm d}=10^5$), the Alfv\'en mode dominates the energy budget with a fraction of $\sim$50\%, nearly independent of wavenumber. The slow mode contributes the second-largest fraction, $\sim$20--40\%, while the fast mode contributes the least, peaking at $\sim$20\% at wavenumbers $k\sim3$--5 and remaining near 10\% at larger $k$.

In the intermediate-coupling regime ($\gamma_{\rm d}=10^3$), where neutral–ion decoupling begins, the slow-mode fraction increases steadily toward large wavenumbers (small scales), while the fractions of the Alfv\'en and fast modes decrease. This trend becomes even more pronounced in the weak-coupling regime ($\gamma_{\rm d}=25$), in which the overall turbulent kinetic energy is strongly damped by ion–neutral collisional friction. The Alfv\'en-mode fraction decreases from $\sim$50\% at $k<10$ to $\sim$10\% at $k>10$. In contrast, the slow-mode fraction rises from $\sim$40\% at $k<10$ to $\sim$60\% at $k$ between 10 and the numerical dissipation scale ($k\sim50$). The fast-mode fraction shows a mild increase toward large wavenumbers but remains below 20\% across the inertial and dissipation ranges.

\begin{figure*}[h]
\centering
\includegraphics[width=0.9\linewidth]{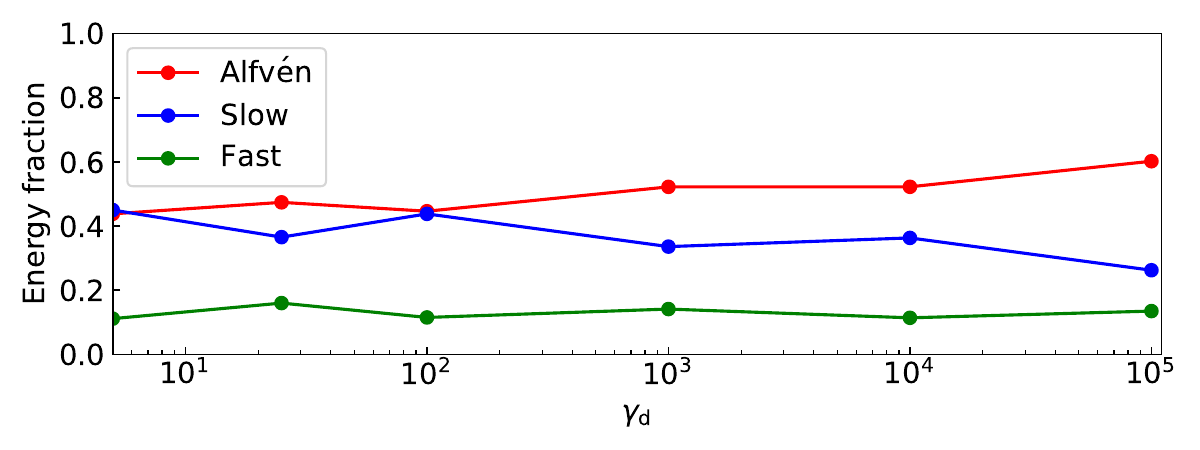}
        \caption{Turbulent kinetic energy fraction in the Alfv\'en (red), slow (blue), and fast (green) modes as a function of the coupling strength, parameterized by the drag coefficient $\gamma_{\rm d}$. $\gamma_{\rm d}=1\times10^5$ represents the strong-coupling case and the decoupling and damping of MHD turbulence gradually happen towards smaller $\gamma_{\rm d}$. Each fraction is defined as the total kinetic energy of each mode normalized by the total kinetic energy of all three modes.}
    \label{fig:fraction}
\end{figure*}

In Fig.~\ref{fig:fraction}, we present the kinetic energy fraction of each MHD mode as a function of the coupling strength, parameterized by the drag coefficient $\gamma_{\rm d}$. Each fraction is defined as the total kinetic energy of a given mode—integrated over all wavenumbers—normalized by the total kinetic energy summed over all three modes. We find that, while the total kinetic energy is reduced in the weak-coupling regime, the kinetic energy fraction of the fast mode remains essentially independent of $\gamma_{\rm d}$ at a level of $\sim$10\%. This indicates that neutral–ion decoupling and damping do not significantly alter the energy partition of the fast mode. In contrast, the Alfv'en-mode fraction decreases from $\sim$60\% at $\gamma_{\rm d}=1\times10^5$ to $\sim$45\% at $\gamma_{\rm d}=5$, while the slow-mode fraction exhibits a corresponding increase from $\sim$30\% to $\sim$45\%. Note that these fractional contributions may also vary if the turbulence driving force is not purely solenoidal or if the plasma $\beta$—and consequently the $M_A$ and $M_s$—changes.

\section{Conclusion}

In this work, we have investigated how ion–neutral collisional damping affects the spectral properties and energy partition of compressible MHD turbulence by analyzing a suite of 3D two-fluid simulations with varying coupling strengths. By decomposing the turbulent velocity field into Alfv\'en, slow, and fast modes, we have characterized how the relative contributions of these modes evolve from the strong-coupling regime to the weakly coupled, partially ionized regime. Our main findings are as follows:
\begin{itemize}
\item In the strong-coupling limit, the Alfv\'en and slow modes follow near-Kolmogorov $k^{-5/3}$ spectra, while fast modes exhibit a steeper spectrum approaching $k^{-2}$. Once neutral–ion damping becomes important, all three modes steepen, with the Alfv\'en mode reaching a dissipation-dominated slope close to $k^{-4}$ and the slow and fast modes showing slightly less spectral steepening.
\item Across all coupling strengths, the Alfv\'en mode carries the largest fraction of the total kinetic energy, contributing roughly 50–60\% in the strong-coupling regime. This confirms that Alfv\'enic fluctuations remain the energetically dominant component of compressible MHD turbulence in both fully and partially ionized conditions.
\item As the coupling weakens, the slow-mode fraction grows significantly toward high wavenumbers. This behavior reflects the resilience of slow-mode fluctuations to neutral–ion damping and highlights their increasing importance in the small-scale fluctuations.
\item  While the total kinetic energy is reduced in the weak coupling regime, the fast mode consistently contributes $\sim$10\% of the total kinetic energy across all coupling strengths. Its fraction remains essentially unchanged from the strong-coupling to the weak-coupling regime, indicating that fast-mode energetics are comparatively insensitive to neutral-ion decoupling and damping.
In contrast, the Alfv\'en mode's energy decreases towards the weak-coupling regime while that of the slow mode increases.
\end{itemize}

\section{Acknowledgments}
Y.H. acknowledges the support for this work provided by NASA through the NASA Hubble Fellowship grant No. HST-HF2-51557.001 awarded by the Space Telescope Science Institute, which is operated by the Association of Universities for Research in Astronomy, Incorporated, under NASA contract NAS5-26555. This work used SDSC Expanse CPU and NCSA Delta CPU through allocations PHY230032, PHY230033, PHY230091, PHY230105,  PHY230178, and PHY240183, from the Advanced Cyberinfrastructure Coordination Ecosystem: Services \& Support (ACCESS) program, which is supported by National Science Foundation grants \#2138259, \#2138286, \#2138307, \#2137603, and \#2138296. 

\bibliography{reference}{}
\bibliographystyle{apalike} 

\end{document}